\documentclass[prd,
 reprint,
 amsmath,amssymb,
 aps,nofootinbib,
 superscriptaddress
]{revtex4-2}
\usepackage{bm}
\PassOptionsToPackage{linktocpage}{hyperref}
\usepackage[hyperindex,breaklinks,hidelinks]{hyperref}
\usepackage{enumitem}
\usepackage{slashed}
\usepackage[dvipsnames]{xcolor}

\renewcommand{\theta}{\vartheta}

\usepackage{subcaption}

\renewcommand{\vec}[1]{\ensuremath{\boldsymbol{#1}}}

\newcommand{\Ll}{\mathcal{L}} 

\usepackage{tikz}
\tikzset{every picture/.style={line width=0.75pt}} 
\usetikzlibrary{shapes.geometric}
\usepackage{orcidlink}
\usepackage{comment}
\usepackage{array}
\usepackage{mathtools}

\usepackage{etoolbox}

\begin{document}

\title{TeV Window to Grand Unification: Higgs's Light Color Triplet Partner}

\author{Gia Dvali} 
\email{gdvali@mpp.mpg.de}
\affiliation{Arnold Sommerfeld Center, Ludwig-Maximilians-Universit\"at, Theresienstra{\ss}e 37, 80333 M\"unchen, Germany}
\affiliation{Max-Planck-Institut f\"ur Physik, F\"ohringer Ring 6, 80805 M\"unchen, Germany}

\author{Otari Sakhelashvili} 
\email{otari.sakhelashvili@sydney.edu.au}
\affiliation{Sydney Consortium for Particle Physics and Cosmology, 
School of Physics, The University of Sydney, NSW 2006, Australia
}

\author{Anja Stuhlfauth\,\orcidlink{0009-0005-0920-379X}} 
\email{anjast@mpp.mpg.de}
\affiliation{Arnold Sommerfeld Center, Ludwig-Maximilians-Universit\"at, Theresienstra{\ss}e 37, 80333 M\"unchen, Germany}
\affiliation{Max-Planck-Institut f\"ur Physik, F\"ohringer Ring 6, 80805 M\"unchen, Germany}

\date{\today}

\begin{abstract} 

The color-triplet partner of the Higgs doublet, called a $T$-particle, is a universal feature of Grand Unification. It has been shown some time ago that this particle can be accessible for direct production in collider experiments. In this paper we point out that the $T$-particle represents a simultaneous low-energy probe of baryon number violation as well as of the origin of the neutrino mass, linking the mediation of proton decay with oscillations of the neutron into a sterile neutrino. We point out a triple correlation between its collider signatures, proton decay measurements and the searches for the magnetic resonance disappearance of free neutrons in cold neutron experiments. In this way, the $T$-particle can provide a diversity of correlated experimental windows into Grand Unification.

\end{abstract}

\maketitle

\section{Introduction}

The goal of the present paper is to show that 
a  light color-triplet partner of the Higgs doublet, 
predicted in \cite{Dvali:1992hc, Dvali:1995hp},  can serve as a low energy window to Grand Unification via correlated signatures
in different categories of experiment.  

Grand Unified Theories (GUTs) provide a symmetry-based framework for the idea of the unification of forces. 
The minimal structure is provided by the $SU(5)$ model of Georgi and Glashow~\cite{Georgi:1974sy} which embeds the $G_{SM} = SU(3)_c\times SU(2)_w \times U(1)_Y$  symmetry of the  Standard Model (SM) in its compact covering group $SU(5)$. 
This gauge symmetry is Higgsed (``broken") down to the SM group at the GUT scale, $M_G$.  The precise value of this scale depends on the low-energy particle content of the theory as well as on the threshold corrections from the physics at the scale $M_G$.
Already in the minimal case, the running of the coupling constants shows that the GUT scale is typically very high, $M_G \sim 10^{15-16}$GeV \cite{Georgi:1974yf}.  This is also in accordance with the phenomenological bounds on proton decay mediated by the $X$ and $Y$ gauge bosons 
corresponding to the broken generators from $SU(5)/G_{SM}$.

The universal feature of Grand Unification is the presence of the color-triplet partner of the SM Higgs doublet, $H$. 
We shall denote this particle by $T$.
Due to the GUT symmetry, the $T$-particle is forced to participate in the $SU(5)$-invariant Yukawa couplings with quarks and leptons along with $H$.
Correspondingly, in the limit of an unbroken $SU(5)$-symmetry, it mediates processes that violate baryon and lepton numbers.
A particularly important process is proton decay
which is subject to very stringent experimental bounds, with a lifetime longer than $\sim 10^{34}$yr.

In order to accommodate such bounds, ordinarily, the $T$-particle is assumed to have a mass $\sim M_G$.  This goes in contrast with the mass of its Higgs partner which is around $\sim 100$ GeV. This mass difference is traditionally referred to as the ``doublet-triplet splitting" problem. 
 
However, this is only a naturalness issue, rather than a consistency one: since the GUT symmetry is broken, the mass splitting can be easily arranged via fine-tuning.
The extensions of the theory can 
allow for more natural solutions either based on symmetry (e.g., within the context of supersymmetric extensions of GUTs) or a cosmological relaxation of the Higgs mass \cite{Dvali:2004tma}. 
In this work, we shall not enter into the naturalness issues of the mass-splitting scenarios which generate a large mass for the $T$-particle. 

From an observational point of view, a universal drawback of such scenarios is that they eliminate any chance of direct detection of the $T$-particle in collider experiments. 
However, an alternative realization of the theory exists in which such detection is possible. 
The idea, originally proposed in \cite{Dvali:1992hc, Dvali:1995hp}, is that, instead of the masses, the splitting can take place among the Yukawa couplings of $H$ and $T$ particles.
As shown in this work, the breaking of GUT symmetry allows for a very strong relative suppression of the Yukawa couplings of the $T$-boson with ordinary quarks and leptons.
This eliminates the necessity of a large mass splitting, 
thereby liberating the $T$-boson to be nearly degenerate 
with the SM Higgs. 

As a result, in this scenario the $T$-boson emerges as a light particle,
with mass $m_T \sim$ TeV, but with suppressed baryon and lepton number violating couplings.    
This allows for the $T$-particle to be potentially accessible at colliders, without conflicting with other phenomenological bounds. 
  
This setup opens up the possibility of a direct detection of GUT's most universal feature in particle collisions as well as the correlation of its collider signatures with other observations.

In the present paper, 
we point out some new implications for baryon and lepton number violating processes involving the neutrino sector.  
In particular, we point out a triple correlation between the signatures
in the following three classes of experiments:  
{\it 1)} a direct production of the $T$-particle at high energy accelerators; 
{\it 2)} the processes mediated by the $T$-boson in proton decay measurements; 
and {\it 3)} the resonant transitions of a free neutron into the sterile neutrino, induced by the $T$-particle, potentially accessible in experiments with cold neutrons.

Before proceeding, we would like to refer to two parallel complementary papers exploring implications of the $T$-particle for flavor physics \cite{Tonginpreparation} and for cosmology \cite{Annainpreparation}.
 We would also like to point to various other implications of the $T$-particle with split couplings
\cite{Dvali:1992hc, Dvali:1995hp, Dvali:1996hs,  Gogoladze:1995sd, Dvali:1997qv, Berezhiani:1998hg, Bajc:2002bv, Rakshit:2003wj, 
Dorsner:2006ye, Belotsky:2008se, Berezhiani:2011uu, Dvali:2020uvd}.

\section{The setup}
For simplicity, we shall work in $SU(5)$ theory 
with minimal field content and a single generation of fermions. 
In $SU(5)$ the SM fermions are arranged in a $\bar{5}$-dimensional and a 10-dimensional representation in the following way:
\begin{equation}
    \bar{5}_F = \begin{pmatrix} d^c_r & d^c_g & d^c_b & e^- & -\nu \end{pmatrix}
\end{equation}

and
\begin{equation}
    10_F = \frac{1}{\sqrt{2}} \begin{pmatrix} 0 & u^c_b & -u^c_g & -u_r & -d_r \\ -u^c_b & 0 & u^c_r & -u_g & -d_g \\ u^c_g& -u^c_r & 0 & -u_b & -d_b \\ u_r & u_g & u_b & 0 & e^+ \\ d_r & d_g & d_b & -e^+ & 0 \end{pmatrix}.
 \end{equation}
Here, the superscript $c$ stands for denoting anti-particles, and $r,g,b$ are the three colors of QCD.
All fermions are written in the left-handed basis. 

In addition, for generating the mass of the neutrino we introduce an $SU(5)$-singlet fermion $\nu^c$, also written in the left-handed basis.  
In the left-right basis, this would play the role of a right-handed (sterile) partner of the active left-handed neutrino $\nu$.

The SM Higgs is embedded in a fundamental representation
\begin{equation}
    5_H = \begin{pmatrix} T_r \\ T_g \\ T_b \\ H^+ \\ H^0 \end{pmatrix},
\end{equation}
where the SM Higgs doublet $H$ sits in the lower two components and a new triplet $T$ in the upper three.
    
As usual, the breaking of the $SU(5)$-symmetry down to the SM group is achieved by a Higgs scalar in the adjoint representation which we denote by $24_H$. 
This field acquires a vacuum expectation value (VEV)
\begin{equation} \label{VEV24}
    \langle 24_H \rangle \propto M_G \text{ diag} \Big(1,1,1,-\frac{3}{2}, -\frac{3}{2} \Big).
\end{equation}
This VEV reduces the symmetry of the vacuum to $G_{SM}$. 
The second stage of the symmetry-breaking is achieved by the VEV of the Higgs doublet $\langle H^0 \rangle = v$,  living in the $5_H$-plet, 
$\langle 5_H \rangle=\begin{pmatrix} 0 & 0 & 0 & 0 & v \end{pmatrix}$. 

This field generates masses of fermions via the 
$SU(5)$-invariant  Yukawa couplings. 
The minimal set of renormalizable couplings that can provide masses to all the fermions, including neutrinos, is given by the following structure:
\begin{eqnarray}
    \Ll_{\rm fermi}  & = &  Y_d \: 5_H^* 10_F \bar{5}_F + Y_u \: 10_F 10_F 5_H +  \nonumber  \\ 
    && + \, Y_{\nu} \: 5_H  \bar{5}_F \nu^c +  m_R \nu^c\nu^c \,, 
\end{eqnarray}
where $Y_d, Y_u, Y_{\nu}$ are the Yukawa couplings.  The structure of $SU(5)$ and Lorentz indices is obvious and is not shown explicitly. 
The first row is responsible for the masses of the charged fermions, whereas the second row generates the mass of the neutrino.
Depending on the values of $m_R$ and $Y_{\nu}$, the 
mass of $\nu$ can be either Dirac or Majorana type. 
In particular, for $m_R \gg Y_{\nu} v$, the Majorana mass of the active neutrino, $m_{\nu} \sim (Y_{\nu} v)^2/m_R$, is generated via the Seesaw mechanism.

Decomposing the above operators in terms of SM particles, we immediately discover that the couplings of $T$ and $H$ are related. 
In such a case, the mass of $T$ must be very large in order not to conflict with the experimental bounds on baryon and lepton number violating processes. 
This is the logic behind the standard assumption of the doublet-triplet mass splitting. 
   
However, notice that the masses of ordinary fermions generated from the above Yukawa couplings are not phenomenologically viable.
In particular, there is a wrong prediction of equal masses for down quarks and electrons~\cite{Ellis:1979fg}.
This rules out the simplest theory.  
  
To obtain viable masses, one necessarily needs to go beyond the minimal structure by adding higher-dimensional operators \cite{Ellis:1979fg}. 
These operators must involve the $24_H$-Higgs. This is necessary 
for communicating the message about the breaking of the $SU(5)$-symmetry to the Yukawa sector.
In a similar spirit, the effective non-renormalizable operators can affect the unification of the gauge couplings \cite{Shafi:1983gz} (for a recent analysis in this direction, see \cite{Senjanovic:2024uzn}).

Now, the idea of \cite{Dvali:1992hc} is that the high-dimensional operators with $24_H$ necessarily undo the $SU(5)$-relations among the couplings of $H$ and $T$. 
As a result, in the low energy theory these couplings appear as independent parameters. 
In particular, the couplings of the color-triplet $T$ with the fermions can be arbitrarily suppressed. 
This liberates the mass of the $T$-particle, allowing it to be within the range of direct experimental accessibility at colliders.

For example, to the lowest order in $24_H$, there exist the following operators that correct the structure of the Yukawa couplings, 	
\begin{equation}\label{eq:5dimoperators}
    \begin{split}
        \Ll_{\text{5-dim}} \supset& \, \frac{g_1}{\Lambda} 5_H^* 10_F 24_H \bar{5}_F +  \frac{g_2}{\Lambda} 5_H^* 24_H 10 _F \bar{5}_F +\\
        +& \, \frac{g_3}{\Lambda} 10_F 10_F 24_H 5_H + \frac{g_4}{\Lambda} 24_H 10_F 10_F 5_H +  \\
        +& \, \frac{g_5}{\Lambda} 5_H 24_H \bar{5}_F \nu^c \,,
    \end{split}
\end{equation}
where $\Lambda$ is a cutoff scale and the $g_i$-s $(i=1,2,...5)$ are numerical coefficients.  
It is clear that upon inserting the VEV of the $24_H$ (\ref{VEV24}) in the above operators, we get the effective low energy theory in which the $SU(5)$ relations among the Yukawa couplings are completely shattered. 

For example, already the addition of the $g_2$-operator  splits the 
$SU(5)$ invariant coupling $5_H^* 10_F \bar{5}_F$ between 
$T$ and $H$ as $Y_d +  
g_2 \frac{M_G}{\Lambda}$  and  $Y_d  - 
g_2 \frac{3M_G}{2\Lambda}$ respectively. Other couplings undergo 
similar splittings. This shows that the effective coupling of the 
$T$-particle can suppress proton decay to the observable level while accommodating the required Yukawa couplings of the Higgs.

In general, one has to include all possible higher-order operators.
Their precise structure depends on the cutoff $\Lambda$ and the specifics of the fundamental theory operating above it.
However, there are some general consistency arguments indicating that the cutoff cannot be arbitrarily far from $M_G$. 

In particular, in any theory propagating $N$ particle species, gravity imposes a universal non-perturbative upper bound on the cutoff scale given by $\Lambda < M_P/\sqrt{N}$, where $M_P$ is the Planck mass~\cite{Dvali:2007hz, Dvali:2007wp}\footnote{This bound originates from the consistency of black hole physics 
and is fully non-perturbative in nature. It is therefore exact and cannot be removed by any resummation of the series. The perturbative arguments supporting it can also be given (e.g., see \cite{Dvali:2001gx, Veneziano:2001ah}).}.
Notice that already within the SM, $N \sim  100$, and in GUTs it is naturally even higher. 
This brings the upper bound on $\Lambda$ at least an order of magnitude below the Planck scale.
This fact is important as it tells us that, already from pure gravity, the corrections to the low-energy couplings cannot be ignored.

In practice, $\Lambda$ can be even lower than the species scale 
and not far from the GUT scale.
In particular, the operators of the type (\ref{eq:5dimoperators}) can be obtained from renormalizable interactions after integrating out multiplets with masses around $M_G$.  
The explicit realizations of such scenarios can be found in~\cite{Dvali:1992hc, Dvali:1995hp}. 
However, specifying the structure of the high-energy theory is beyond the scope of the present paper. 
Instead, we shall treat the Yukawa couplings of the $T$-particle as free parameters of the low energy theory, and study certain interesting ranges of the parameter space.

With all generality, we parameterize the interactions of the $T$-particle and the ordinary Higgs doublet, $H$, with the SM fermions by the following effective Lagrangian:
\begin{equation} \label{TandH}
    \begin{split}
        \Ll \supset & \: g_{Tud} \: T^* \, u^c d^c \; + g_{TQL} \: T^*\,  Q L \;   +   \\ 
        +&  g_{Tue} \: T \, u^c e^c  \;+ g_{TQQ} \:  T \, Q Q \; +  g_{Td\nu} \: T \, d^c \nu^c + \\  
        +& Y_d' \:  H^* \, Q d^c \; + Y_e' \:  H^* \, L e^c \; + Y_u' \: H \,  Q u^c\; +  \\
        +&  Y_{\nu}' \: H \, L \nu^c.       
    \end{split}
\end{equation}
In the limit of an exact $SU(5)$-symmetry, we would have $g_{Tud} = g_{TQL}  =  Y_d' = Y_e' $, $g_{Tue} = g_{TQQ} = Y_u' $ and $g_{Td\nu}  = Y_{\nu}'$. 
However, through high-dimensional operators of the type (\ref{eq:5dimoperators}), the VEV of $24_H$ splits them arbitrarily. 
    
The couplings of the doublet are fixed by the Higgs VEV ($v$) and the masses of corresponding fermions in the usual way:  $Y_d' = m_d/v,  Y_e'= m_e/v,  Y_u' = m_u/v$. 
The analogous relation for the neutrino coupling/mass depends on the value of $m_R$ and will be discussed below.  
     
The couplings of the triplet shall be constrained by various processes with baryon and lepton number violation.

\section{ 
\texorpdfstring{
$T$}{Lg}-Phenomenology} 

\subsection{Proton and Neutron decays}

The exchange of the $T$-particle can mediate baryon and lepton number violating processes. 
Let us first consider the processes of baryon (proton or neutron) decay.   
Such decays require an exchange that on one hand converts a quark into a quark and on the other hand a quark into a lepton. 
Correspondingly, the amplitudes of such processes are governed by the different cross products of the couplings of the type $T$-quark-quark ($g_{Tud},   g_{TQQ}$)  and $T$-quark-lepton ($g_{Tue}, g_{TQL}, g_{Td\nu}$). 

\begin{figure}[t]
    \centering
    \includegraphics[width=0.384\textwidth]{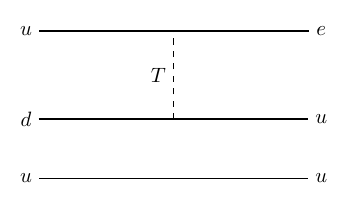}
    \vspace{-2mm}
    \caption{An example of a proton decay process mediated  via the $T$-particle.}
    \label{fig:protondecay}
\end{figure}

For a clearer visualization, it is useful to write down the set of corresponding baryon and lepton number violating effective four-fermion operators obtained by the tree-level exchanges of the $T$-particle,
\begin{equation} \label{4fermi}
    \begin{split}
        \Ll_{eff} \supset & \: \frac{g_{TQQ}g_{Tue}}{m_T^2} \:  Q Q u^{c*}e^{c*} \; + \; 
        \frac{g_{Tud}g_{Tue}}{m^2_T} \: u^cd^c u^ce^c \;  +  \\ 
        +&  \, \frac{g_{Tud}g_{Td\nu}}{m_T^2} \:  u^c d^c d^c \nu^c  \;
        + \frac{g_{TQQ}g_{TQL}}{m_T^2} \:  Q Q Q L \; +
          \\  
        +& \, \frac{g_{Tud}g_{TQL}}{m^2_T} \:  u^c d^c \, Q^*L^* \;  
         + \; \frac{g_{TQQ}g_{Td \nu}}{m_T^2}  \: Q Q d^{c*}\nu^{c*} \,,
    \end{split}
\end{equation}   
where $m_T$ is the mass of the $T$-particle and the star stands for conjugation. 
These operators can mediate proton and neutron decay in various channels.   
For example, consider a proton decay via the channel $p \rightarrow e^+ \pi^0$.
A representative exchange diagram is given in \autoref{fig:protondecay}.

Taking the mass of the triplet around its current collider bound, $m_T \sim$ TeV, the experimental bound on the proton lifetime, $\tau_p > 10^{34}$yr~\cite{Super-Kamiokande:2020wjk}, implies the following constraint on the cross products of $T$-quark-quark and $T$-quark-lepton couplings, 
\begin{equation} \label{gg}
   {\rm (T-quark-quark) \cdot (T-quark-lepton)}  < 10^{-26}\,,
\end{equation}
where  the product runs over $(g_{Tud} \cdot g_{Tue}), (g_{TQQ} \cdot g_{TQL}), (g_{Tud} \cdot g_{TQL}), (g_{TQQ} \cdot g_{Tue})$.
\footnote{We note that throughout the paper the displayed bounds on the parameters of the theory must be understood as order of magnitude wise relations with numerical factors not shown explicitly.}

Other processes of the proton decay into a charged lepton,
such as $p \rightarrow e^+ \gamma$, give weaker bounds. 

Likewise, the above operators generate the process of neutron decay
into a charged lepton
$n \rightarrow e^+ \pi^-$. These are within experimental bounds provided the couplings satisfy the proton decay constraints (\ref{gg}).

We now turn to the processes that involve the neutrino sector. 
For this let us first identify the neutrino mass eigenstates.
The neutrino mass originates from the following Yukawa coupling with $H$,
\begin{equation}    
   Y_{\nu}' \: H \, L \nu^c \,+ \,  m_R \nu^c \nu^c  \,.     
\end{equation}
For a non-zero Majorana mass, $m_R$, the $\nu^c$ entering this coupling is not a mass eigenstate. 
Instead, it is a superposition of the mass eigenstates $\nu^{c \prime}$ and $\nu'$ of the sterile and active neutrino flavors. 
For $m_R \gg Y_{\nu}' v$,  the mixing angle can be approximated as  $\theta \simeq   Y_{\nu}' v/m_R$ and can be expressed via mass eigenvalues as $\theta \simeq \sqrt{m_{\nu}/m_R}$, where $m_{\nu} \simeq  (Y_{\nu}' v)^2/m_R$ is the mass of the active SM neutrino.
The interaction and mass eigenstates are therefore related as 
 \begin{equation}  \label{nus}  
     \nu^c \, \simeq \,  \nu^{c \prime} + \sqrt{\frac{m_{\nu}}{m_R}} 
   \nu'\,, ~~ 
     \nu  \, \simeq \,  \nu'  -  \sqrt{\frac{m_{\nu}}{m_R}} 
   \nu^{c \prime}\,.
 \end{equation}

 Next let us turn to the couplings of the $T$-particle that can 
 mediate proton decay with a neutrino in the final state: 
\begin{equation} \label{TProton}
    \begin{split}
        & g_{TQQ} \:  T \, Q Q \; + g_{TQL} \: T^*\,  Q L \;   +  \\ 
        +& g_{Tud} \: T^* \, u^c d^c \; + g_{Td\nu} \: T \, d^c \nu^c .        \end{split}
\end{equation}
The corresponding four-fermion operators obtained by $T$-exchange come from the second and third rows of (\ref{4fermi}). 
For clarity, we shall present them here explicitly in terms of their neutrino entries, 
 \begin{equation} \label{4Nu}
    \begin{split}
        &  \, \frac{g_{Tud}g_{Td\nu}}{m_T^2} \:  u^c d^c d^c \nu^c  \;
        + \frac{g_{TQQ}g_{TQL}}{m_T^2} \:  u d d \nu \; +
          \\  
        +& \, \frac{g_{Tud}g_{TQL}}{m^2_T} \:  u^c d^c \, d^*\nu^* \;  
         + \; \frac{g_{TQQ}g_{Td\nu}}{m_T^2}  \: u d d^{c*}\nu^{c*} \,.
    \end{split}
\end{equation}   
The participation of the neutrino mass eigenstates in these couplings goes via the substitutions according to the equation (\ref{nus}). 
 
For $m_R > m_p$, the sterile neutrino participates in the proton decay only as a virtual state.
We shall focus on such a case. 
The couplings in (\ref{TProton}) and the corresponding effective operators (\ref{4Nu}) 
lead to the processes of proton decay in the channel $p \rightarrow \pi^{+}\nu'$.

\begin{figure}[t]
\begin{subfigure}{0.48\textwidth}
    \includegraphics[width=0.8\textwidth]{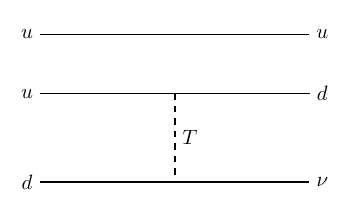}
    \vspace{-3mm}
    \caption{via $g_{TQL}$}
    \label{fig:protondecaytonuL2}
\end{subfigure}
\begin{subfigure}{0.48\textwidth}
    \includegraphics[width=0.8\textwidth]{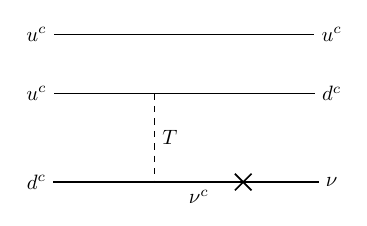}
    \vspace{-2mm}
    \caption{via $g_{T d \nu}$}
    \label{fig:protondecaytonuL}
\end{subfigure}
\caption{ Examples of diagrams responsible for proton decay into a pion and an active neutrino via different couplings.}
\end{figure}

Some examples of representative diagrams are given in 
\autoref{fig:protondecaytonuL2} and \autoref{fig:protondecaytonuL}. 
 The effective operators in (\ref{4Nu}) involving the coupling $g_{TQL}$
 lead to the following lifetimes of the proton,
 \begin{equation} \label{rateL}
   \tau_{p}' \, \sim  \,     
  C \, \frac{m_T^4}{m_p^5} \,,
\end{equation}
 where the parameter $C$ is set by the coefficient 
 of the corresponding operator as $C = (g_{TQQ} \,g_{TQL})^{-2}$
 or $C = (g_{Tud} \,g_{TQL})^{-2}$. 
 Analogously, the operators involving $g_{Td\nu}$ give the lifetimes,
\begin{equation} \label{rateR}
   \tau_{p} \, \sim  \,  C\, \frac{m_R}{m_{\nu}} \, \frac{m_T^4}{m_p^5}\,,
\end{equation}
with $C = (g_{TQQ} \,g_{Td\nu})^{-2}$
 or $C = (g_{Tud} \,g_{Td\nu})^{-2}$.

Here we distinguish the would-be lifetimes by a prime in order to differentiate between the processes involving the 
couplings $g_{TQL}$ and $g_{Td\nu}$. 
Obviously, to have a neutrino in the final state, 
one of these couplings is necessary.  
The processes that go through $g_{Td\nu}$ exhibit an extra enhancement of the proton lifetime by a factor $m_R/m_{\nu}$ which comes from the mixing angle in (\ref{nus}). 

The experimental bound on the proton lifetime in the decay channel  $p \rightarrow \pi^{+} \nu'$, $\tau_p > 10^{32}$yr~\cite{Super-Kamiokande:2013rwg}, translates into the following bounds on the coefficients in (\ref{rateL}) and (\ref{rateR}) respectively,   
\begin{equation} \label{Cbounds}
    C\,  > 10^{50}\,~~ {\rm and} ~~ \, C\,\frac{m_R}{m_{\nu}} > 10^{50}\,.
\end{equation}

Notice that if we assume the $SU(5)$-relation, $g_{Td\nu} = Y_{\nu}'$, in the expression (\ref{rateR}) the coupling and $m_R$ can be traded for the known parameters of the SM such as the Higgs VEV, $v$, and the neutrino mass $m_{\nu}$. 
This gives 
\begin{equation} 
    \tau_{p} \sim (C g_{Td\nu}^{2}) 
    \left (\frac{v}{m_{\nu}}  \right )^2    
    \frac{m_T^4}{m_p^5}\,.
\end{equation}
Taking $m_{\nu} \sim 10^{-2}$eV, the experimental bound $\tau_p > 10^{32}$yr~\cite{Super-Kamiokande:2013rwg} implies 
\begin{equation}
    g_{Tud}\,, \, g_{TQQ} < 10^{-12} \,.
\end{equation}
The above constraints also satisfy the bounds on neutron decay. 

\begin{figure}[ht]
    \centering
    \includegraphics[width=0.384\textwidth]{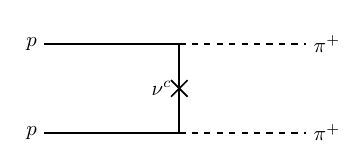}
    \vspace{-3mm}
    \caption{Double proton decay}
    \label{fig:doubleprotondecay}
\end{figure}

The Majorana mass of $\nu^c$ can also lead to a process of double proton decay $pp \rightarrow \pi^+ \pi^+$ (see \autoref{fig:doubleprotondecay}).
However, the experimental bound on this process is $10^{31}$yr~\cite{Super-Kamiokande:2015jbb}, which puts weaker bounds on our parameters as compared to the bounds discussed above.

\subsection{Neutron oscillations into the sterile neutrino}  
 
We now wish to point out that the light $T$-particle can induce the interesting process of neutron oscillations into the sterile neutrino. 
 This is possible if $m_R$ is close to the mass of the neutron $m_n$.
The triplet exchange generates a mixing mass term between $n$ and $\nu^{c \prime}$, which we shall denote by $m_{n\nu^c}$.
This mixing term induces the oscillations of a free neutron into the sterile neutrino with the period $\tau_{n\nu^c} = m_{n\nu^c}^{-1}$. 
The remarkable thing is that the transition process is correlated with the process of proton decay in the channel $p \rightarrow \pi^{+} \nu'$, essentially with no free parameters
involved. 

\begin{figure}[ht]
\begin{subfigure}{0.48\textwidth}
    \includegraphics[width=0.8\textwidth]{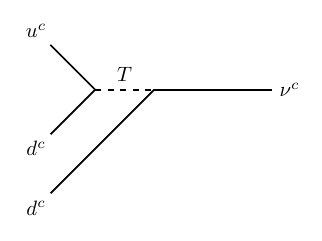}
    \vspace{-5mm}
    \caption{via $g_{Td \nu}$}
    \label{fig:sterileneutrinoosc}
\end{subfigure}
\begin{subfigure}{0.48\textwidth}
    \includegraphics[width=0.75\textwidth]{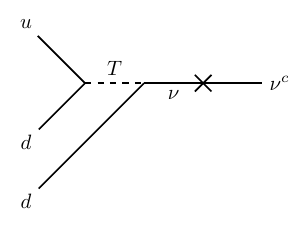}
    \vspace{-5mm}
    \caption{via $g_{TQL}$}
    \label{fig:sterileneutrinoosc2}
    \end{subfigure}
\caption{Examples of diagrams mixing neutron with sterile neutrino via different couplings. }
\label{fig:bothsteileneutrinoosc}
\end{figure}

Indeed, all processes leading to this decay have counterparts 
that generate the mixing term $m_{n\nu^c}$. 
For example, a pair of corresponding representative diagrams is depicted in \autoref{fig:bothsteileneutrinoosc}.

The induced oscillation periods are in one-to-one correspondence with the expressions for the proton lifetimes.
The difference is that the processes that go through the coupling $g_{TQL}$  exhibit a relative enhancement by a factor $\sqrt{\frac{m_R}{m_{\nu}}}$ in neutron oscillation period versus the proton lifetime, as compared to the processes that go via  $g_{Td\nu}$.

To be more explicit, for processes mediated via $g_{TQL}$, we have the following relation between the two time-scales,
\begin{equation}\label{tauL}
    \tau_{n\nu^c} \simeq   
    \sqrt{\frac{\tau_p'}{m_{\nu}}}\,,
\end{equation}
whereas for the processes mediated via $g_{Td\nu}$, the analogous relation is 
\begin{equation}\label{tauR}
    \tau_{n\nu^c}    
    \simeq  m_n^{-1} \sqrt{\tau_p m_{\nu}}\,.  
\end{equation}  
In both expressions we took $m_R = m_n = m_p$.
It is clear that for $\tau_p'$ and $\tau_p$  at the current experimental bound on the proton lifetime, the expression (\ref{tauR}) gives a shorter oscillation time. 
Taking $\tau_p \sim 10^{32}$yr~\cite{Super-Kamiokande:2013rwg}, we get for the oscillation period of a free neutron into the sterile neutrino $\tau_{n\nu^c} \sim 10^{2}-10^{3}$s.

This time-scale is within the range
of interest for the experiments that measure the disappearance of cold neutrons into hidden species \cite{nEDM:2020ekj,Ban:2023cja}.
Both experiments are set up for testing the transition of a neutron into a nearly degenerate fermion. Therefore, they are ready-made for probing the proposed transition of the neutron into the sterile neutrino. 
These experiments place the ultracold neutrons in a magnetic field $B$ which changes the energy of the neutron proportional to its magnetic moment $\mu_n$, $\Delta E = \mu_n B$.
When the energies of the neutron and the partner particle are nearly degenerate, a resonance 
increase in the neutron disappearance probability takes place.
Ban et al.~\cite{Ban:2023cja} performed a scanning of the magnetic field in steps of $\Delta B=3 \mu$T within the range of $50 - 1100 \mu$T.
Correspondingly, they were able to put a bound on the neutron disappearance time $\tau_{n n'} > 1$s for a mass splitting $2-69$peV. 
In the present context this bound translates as the corresponding constraint on  $\tau_{n\nu^c}$  for the ``resonant" regime $m_n + \Delta E \simeq m_R$. 

If the splitting between $m_n$ and $m_R$ is significant, the transitions of the free neutron into $\nu^c$ are suppressed. 
However, for certain values $m_n >  m_R$, the mass $m_R$ can come in resonance with some energy levels $E_n$ of a bound neutron inside the nucleus. 
In this case, the bounds on the couplings of $T$ become more stringent in order to accommodate the neutron disappearance bounds from the nuclei, 
which gives the time-scale $> 10^{30}$yr \cite{SNO:2022trz,KamLAND:2005pen}.

   Notice that for having a resonant transition, 
  it is sufficient that  the difference between $m_R$ and the neutron energy level $E_n$ is within the binding energy of the neutron $\Delta E_n \equiv m_n - E_n$ which is $\Delta E_n \sim$ MeV. The reason is that in such a case the neutron can transition into a sterile neutrino with a momentum $\vec{p}$ such that $E_n = \sqrt{m_R^2 + |\vec{p}|^2}$.  This is because, for $|\vec{p}| <  \sqrt{\Delta E_n m_R}$
  the momentum can be absorbed by the recoil of the nucleus.
  
 Since the would-be oscillation period is much longer than 
 the nuclear de-excitation time as well as the escape time of $\nu^c$, 
 the inverse transition is not possible. Correspondingly, the outcome of 
 the process will be a transition of the ``host"  atom into an isotope 
 with fewer neutrons.  This transition will be accompanied by the emission of photons of nuclear energy.  
  The non-observation of this effect puts the bound 
  on the disappearance time $\sim 10^{30}$yr \cite{SNO:2022trz,KamLAND:2005pen}.
  
  Notice that once the resonant $n-\nu^c$ transition for  a nuclear neutron satisfies this bound, the corresponding two-body decays
  of the neutron and the proton, such as $p \rightarrow \pi^{+} \nu$ and 
  $n \rightarrow \pi^{0} \nu$, are well below the experimental limits.

To summarize, for various resonant values of $m_R$ with neutron energy levels, the $T$-particle can mediate transitions between the neutron and $\nu^c$ for a free as well as for a bound neutron.  This 
provides a new set of low energy probes of Grand Unification linked 
with the neutrino mass. In principle, the resonant values can be scanned by the external magnetic field. 
One can say that in this setup the neutron energy levels provide a spectroscopy of the sterile neutrino.

Of course, in this setup we  treat the mass of the sterile neutrino as 
a parameter of the effective theory and derive its observable consequences.  
The exciting feature is that it gives a chance to probe GUT physics in low energy measurements with some spectacular signatures and correlate it with other observations.

\subsection{Neutron-anti-neutron transitions} 
   
\begin{figure}[ht]
    \centering
    \includegraphics[width=0.384\textwidth]{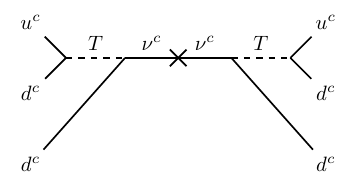}
    \caption{ An example of a diagram generating the neutron-antineutron transition.}
    \label{fig:neutron antineutron}
\end{figure}

Since in the above parameter space $\nu^c$ is a Majorana fermion, its mixing with the neutron automatically induces a mixing between the neutron and the anti-neutron, see \autoref{fig:neutron antineutron}.  
The corresponding transition time for a free neutron is
\begin{equation} \label{tnnbar}
    \tau_{n\bar{n}}  \sim \tau_{n \nu^c}^2 m_n \sim 10^{30} {\rm s} \,, 
\end{equation}
which is very far from the latest experimental limits, 
$\tau_{n\bar{n}} > 0.86 \times 10^{8}$s \cite{Baldo-Ceolin:1994hzw}. 

For a neutron that is in a bound state inside a nucleus, taking into account the additional 
suppression due to a level splitting between $n$ and $\bar{n}$, this 
gives a transition time $\tau_{n\bar{n}}^{(nucl)} \sim 
10^{75}$yr which by many orders of magnitude exceeds 
the current experimental bound $\tau_{n\bar{n}}^{(nucl)} \sim 3.6 \times 
10^{32}$yr \cite{Super-Kamiokande:2020bov}. 

Thus, for the ``resonant" regime $m_n \simeq m_R$, the dominant process allowed by the proton stability constraint is the oscillation between a neutron and a sterile neutrino \footnote{The correlation with proton decay is an important constraint on $n-\bar{n}$ versus $n-\nu^c$ transitions exhibited by the present GUT framework. This goes in sharp difference with alternative extensions of the SM such as the $n-\bar{n}$ oscillations via mirror copies of the neutron \cite{Berezhiani:2020vbe} where such correlations are absent.}. 
In general, in this regime the $n-\bar{n}$ oscillation period $\tau_{n\bar{n}}$ is additionally enhanced by the factor $\tau_{n\nu^c} m_n$ relative to $\tau_{n\nu^c}$. 
This is because the $n-\bar{n}$ transition has to go via an intermediate $\nu^c$ state.

\subsection{Correlations with \texorpdfstring{$T$}{Lg}-production at LHC}

The $T$-particle can be pair-produced at the LHC in the collision of ordinary hadrons. Early discussions of its production mechanisms 
and collider signatures were given in \cite{Dvali:1992hc, Dvali:1995hp, Barbieri:1992yy,
Dvali:1997qv, Belotsky:2008se, Cheung:2002uz}.  
More recent studies of bounds on new colored particles 
in the light of LHC data ~\cite{ATLAS:2022pib,ATLAS:2021mdj,ATLAS:2019gqq,CMS:2017kku,CMS:2016kce,ATLAS:2016onr}  set the lower bound on the mass
of the $T$-particle around TeV.

For this mass, the correlations between various baryon number violating processes were discussed above. 
These processes constrain the lifetime of $T$, which is important for its LHC phenomenology. 

For example,  assuming that the dominant couplings are given by the second row of (\ref{TProton}), the $T$-triplet can decay either into an up and a down quark or into a down quark and a sterile neutrino.
The dominance of the channel depends on the relative strength of the couplings, $g_{Tud}$ versus $g_{Td \nu}$.
In the simplest case, in which the couplings of the doublet and triplet to the sterile neutrino are given by the $SU(5)$-relation, we can estimate, 
\begin{equation}
    g_{Td \nu} \sim \frac{\sqrt{m_\nu m_R}}{v} \sim 10^{-8}.
\end{equation}
The $T$-quark-quark coupling is then constrained by the proton decay as,
\begin{equation}
    g_{Tud} < 10^{-12} \,.
\end{equation}
The corresponding lifetime of the $T$-particle is 
\begin{equation}
    \tau_T \sim 10^{-11} {\rm s} \sim  1 {\rm mm}\,.
\end{equation}
Since this time-scale is much longer than the QCD length, the $T$-triplet will have plenty of time to hadronize and to form a color-singlet state.
In particular, it can form a ``$T$-baryon" state with two fermions of the same color ($Tqq, T^*q^cq^c$), or a ``$T$-meson" state with a fermion of an anti-color ($Tq^c, T^*q$).  
Notice that, since the $T$-particle is a scalar, with such a definition, the $T$-baryon is a boson, whereas $T$-meson is a fermion.

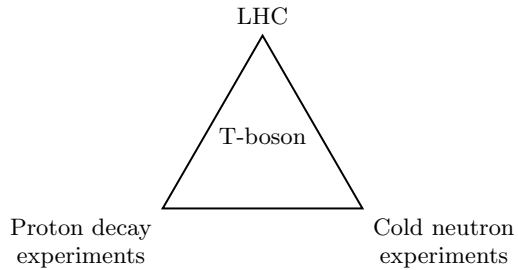
\begin{figure}[ht]
    \centering
\begin{tikzpicture}

\node[isosceles triangle,
	isosceles triangle apex angle=60,
    rotate=90,
	draw,
    fill=none,
	minimum size =2.3cm] (triangle)at (3,0){};

\node[anchor=south] at (triangle.center) {T-boson};
\node[anchor=south] at (triangle.apex) {LHC};
\node[anchor=north east, align=center] at (triangle.left corner) {Proton decay\\ experiments};
\node[anchor=north west, align=center] at (triangle.right corner) {Cold neutron\\ experiments};

\end{tikzpicture}
\caption{A schematic representation of correlations between various experimental searches. }
\label{fig:correlations}
\end{figure}

Assuming that upon its production the $T$-particle is moderately relativistic, the hadronized $T$-particle will decay via a displaced vertex. 
Depending on the precise values of the parameters, the decay can take place within the detector. 
For example, for a $T$-meson the decay products can be an ordinary QCD meson and a sterile neutrino, e.g., 
\begin{equation}
    (Td^c) \rightarrow \pi^0 + \nu^c \,. 
\end{equation}
Analogously, for a $T$-baryon the decay can take place into an ordinary baryon and a sterile neutrino, e.g., 
\begin{equation}
    (T^*u^cd^c) \rightarrow \bar{n} + \nu^c \,. 
\end{equation}
Of course, for more suppressed values of its couplings, the decay can take place outside of the detector, in which case the $T$-hadron will appear as a stable particle. 
The various parameter regimes require more precision studies. 
However, the general message is that for the light GUT color-triplet scenario \cite{Dvali:1992hc, Dvali:1995hp}, the searches for proton decay on one hand, and the cold neutron oscillations/disappearance experiments on the other hand become correlated with each other as well as with the production of a long-lived $T$-hadron states in colliders
   (see \autoref{fig:correlations}).

\section{Conclusion}
    
The color-triplet partner of the Higgs doublet is a model-independent property of any GUT. Therefore, an experimental detection of the $T$-boson would be direct evidence of Grand Unification. 
        
As originally pointed out in \cite{Dvali:1992hc, Dvali:1995hp}, the $T$-particle can be part of the low energy spectrum of the theory with the mass potentially accessible at collider experiments, such as LHC. 
The reason is that already the lowest order operators involving the VEV of $24_H$ are capable of suppressing the baryon and lepton number violating couplings of the $T$-boson to the levels compatible with the current experimental bounds. 
   
This opens up an avenue for simultaneous searches of the signatures of the $T$-particle at high energy colliders as well as in the experiments looking for baryon and lepton number violating low energy processes. 
  
In particular, in the present paper we have pointed out the potential correlations among the searches at three very different sets of experiments.  
The first one comes from the link between proton decay and the oscillations of a cold neutron into a sterile neutrino. 
Namely, the current bounds on the proton lifetime put the $T$-mediated neutron-to-sterile neutrino oscillation processes within the ballpark of sensitivity of cold neutron experiments \cite{Ban:2023cja, nEDM:2020ekj}. 
    
In general, due to its vanishing SM gauge charges, the neutron represents an interesting probe of feebly-coupled new physics.
The early examples of such effects include the neutron transitions into its hidden copies  \cite{Dvali:1999gf, Berezhiani:2005hv, Dvali:2009ne}. 
It was shown recently \cite{Dvali:2023zww} that within the framework of large extra dimensions \cite{Arkani-Hamed:1998jmv}, the ``disappearance" or oscillations of a cold neutron into the extra dimensions is a rather generic phenomenon. 
Moreover,  this effect can be linked with the extra-dimensional origin of the neutrino mass \cite{Arkani-Hamed:1998wuz, Dvali:1999cn}.

The analysis of the present work shows that a similar phenomenon can take place in GUTs. 
Moreover, here too the cold neutron transition is linked with the origin of the neutrino mass.     
Namely, via $T$-exchange a neutron can oscillate into a sterile neutrino which generates the Majorana mass for the SM neutrino. 
Thus, the GUT framework with a light $T$-particle can correlate the $T$-mediated proton and neutron decay processes with the oscillations of cold neutrons into the sterile neutrino species. 
In this way, the cold neutron experiments can potentially turn out to be simultaneous probes of Grand Unification and of  the neutrino mass. 

In addition, the parameter space predicting the signatures for both the above types of measurements, at the same time, is linked with the direct production of the $T$-particle at the LHC.
Thus, the $T$-boson, which represents an intrinsic part of any GUT, can manifest itself in the correlated signatures in three different types of experiments.

Although we have illustrated the effect within the minimal $SU(5)$ 
with a single generation of fermions, it is straightforward to see that it persists in the generalizations to higher GUTs such as $SO(10)$ or $SU(6)$, as well as, in their supersymmetric extensions which have been formulated previously \cite{Dvali:1992hc, Dvali:1995hp, Dvali:1996hs, Berezhiani:2011uu, Dvali:2020uvd}. Moreover, the inclusion of 
couplings of the $T$-particle with heavier generations 
of fermions provides additional baryon decay channels and correlated  transitions of the neutron into the corresponding flavors 
of the sterile neutrinos. 
\\

{\bf Acknowledgments} \\

 It is a pleasure to thank Zurab Berezhiani, Allen Caldwell, Manuel Ettengruber, Anna Jankowsky, 
 Giorgos Karananas and Tong Zhang for discussions on various features of the $T$-particle and of neutron transitions. 
 We thank Stefan Kluth and Sandra Kortner for the discussions on collider signatures. 
This work was supported in part by the Humboldt Foundation under Humboldt Professorship Award, by the European Research Council Gravities Horizon Grant AO number: 850 173-6,
by the Deutsche Forschungsgemeinschaft (DFG, German Research Foundation) under Germany's Excellence Strategy - EXC-2111 - 390814868, and Germany's Excellence Strategy under Excellence Cluster Origins. \\
 
 Disclaimer: Funded by the European Union. Views
and opinions expressed are however those of the authors
only and do not necessarily reflect those of the European Union or European Research Council. Neither the
European Union nor the granting authority can be held
responsible for them.

    \bibliographystyle{utphys}
	\bibliography{bib.bib}

\end{document}